\documentclass[%
 reprint,
]{revtex4-1}

\usepackage{graphicx}
\usepackage{dcolumn}
\usepackage{bm}
\usepackage{color}
\usepackage{braket}
\usepackage{amsmath}
\usepackage{booktabs}
\usepackage{caption}
\usepackage{lipsum}
\usepackage{boxhandler}
\usepackage{lineno}

\def\SIO{Sr$_{2}$IrO$_{4}$}

\begin{document}



\title{Single laser pulse driven thermal limit of the quasi-two dimensional magnetic ordering in \SIO}

\author{Ruitang Wang$^{1,2,3}$}
\author{J. Sun$^{1}$}
\author{D. Meyers$^{4,5}$}
\author{J. Q. Lin$^{1,2,3}$}
\author{J. Yang$^{6}$}
\author{G. Li$^{1}$}
\author{H. Ding$^{2,3}$}
\author{Anthony D. DiChiara$^{7}$}
\author{Y. Cao$^{8}$}
\author{J. Liu$^{6}$}
\author{M. P. M. Dean$^{4}$}
\author{Haidan Wen$^{7,8}$}
\author{X. Liu$^{1}$} 
\email{liuxr@shanghaitech.edu.cn}
\affiliation{$^{1}$School of Physical Science and Technology, ShanghaiTech University, Shanghai 201210, China.}
\affiliation{$^{2}$
Beijing National Laboratory for Condensed Matter Physics and Institute of Physics, Chinese Academy of Sciences, Beijing 100190, China}
\affiliation{$^{3}$
University of Chinese Academy of Sciences, Beijing 100049, China}
\affiliation{$^{4}$
Condensed Matter Physics and Materials Science Department, Brookhaven National Laboratory, Upton, New York 11973, USA.}
\affiliation{$^{5}$
Department of Physics, Oklahoma State University, Stillwater, Oklahoma 74078, USA.}
\affiliation{$^{6}$
Department of Physics and Astronomy, University of Tennessee, Knoxville, Tennessee 37996, USA.}
\affiliation{$^{7}$
Advanced Photon Source, Argonne National Laboratory, Argonne, IL, 60439, USA.}
\affiliation{$^{8}$
Materials Science Division, Argonne National Laboratory, Argonne, Illinois, 60439, USA.}

\begin{abstract}

Upon femtosecond-laser stimulation, generally materials are expected to recover back to their thermal-equilibrium conditions, with only a few exceptions reported. Here we demonstrate that deviation from the thermal-equilibrium pathway can be induced in canonical 3D antiferromagnetically (AFM) ordered \SIO{} by a single 100-fs-laser pulse, appearing as losing long-range magnetic correlation along one direction into a glassy condition. We further discover a `critical-threshold ordering' behavior for fluence above approximately 12 mJ/cm$^2$ which we show corresponds to the smallest thermodynamically stable $c$-axis correlation length needed to maintain long-range quasi-two-dimensional AFM order. We suggest that this behavior arises from the crystalline anisotropy of the magnetic-exchange parameters in \SIO{}, whose strengths are associated with distinctly different timescales. As a result, they play out very differently in the ultrafast recovery processes, compared with the thermal equilibrium evolution. Thus, our observations are expected to be relevant to a wide range of problems in the nonequilibrium behavior of low-dimensional magnets and other related ordering phenomena.
\end{abstract}


\pacs{Valid PACS appear here}

\maketitle

\section{\label{sec:level1}Introduction}
Understanding the mechanisms behind laser manipulation of magnetism in materials is indispensable to many problems in both fundamental and applied science \cite{Beaurepaire1996, RMP2010,KimelNRM2019,  Bigot2009, Nemec2018}. Phenomenologically, the classic three-temperature model\cite{Beaurepaire1996} works quite well in explaining the experimental observations on the evolution of the spins upon external femtosecond-laser stimuli in many systems\cite{RMP2010, Beaurepaire1996, QZhang2006, Kimling2014, Koopmans2010}. In this model, the spin, electron and lattice baths are coupled through a set of mutual interaction constants which govern the energy-flow rates when the system is stimulated into nonthermal equilibrium conditions. Eventually, the spin sector thermalizes back to equilibrium condition after the externally deposited energy equilibrates among the charge, spin and lattice reservoirs\cite{Beaurepaire1996}. Here we report an exceptional observation of single-laser-pulse-induced 
anisotropic partial recovery of the magnetic ordering in \SIO{}, leading to permanent reduction of the interplane magnetic correlation. We propose that this unusual recovery behavior could be explained by the noncooperation of different terms of the exchange interactions dictating the spin ordering. When these exchanges differ significantly in strength, they are unfolded along the time axis\cite{Peter2019,LCLSreview} into different time scales in the ultrafast recovery process. Because of this `time-window mismatch', the weaker exchange terms could be quenched, leading to a deviation from the quasiequilibrium relaxation pathway.

\begin{figure*}
\includegraphics[width=1.0\textwidth]{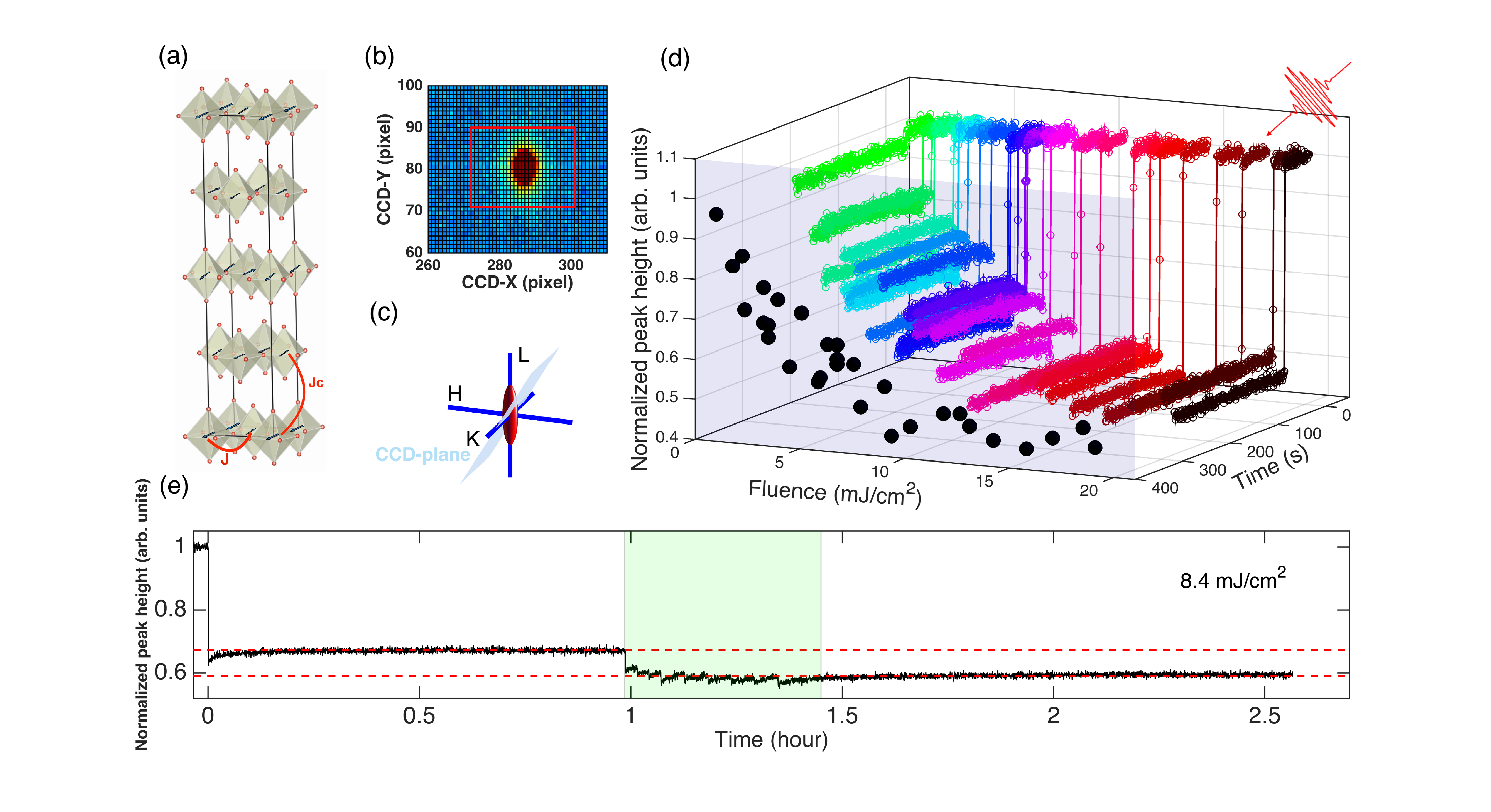}
\captionsetup{justification=raggedright,singlelinecheck=false}
\caption{\label{fig:wide}Single-laser-pulse-induced suppression of the (1 0 16) magnetic reflection. (a) The crystal and magnetic structure of \SIO. J and J$_{c}$ denote the in-plane and interplane exchange couplings, respectively. (b) CCD image of the magnetic peak. (c) Cartoon demonstration of the projection of the CCD plane in the reciprocal space. (d) Response of the 
magnetic Bragg-peak height to a single laser pulse (arrives at t=0) as a function of laser fluence. The black dots are the average of the last 100 s of each curve. (e) Multiple-shot evolution of the magnetic-peak height with moderate laser fluence. Multiple shots arrive in the shaded region and each creates a small intensity drop until the decrease is fully compensated by an initial recovery.}
\label{CCD}
\end{figure*}

\SIO{} is a layered Mott insulator with each Ir site hosting a 1/2 pseudospin\cite{BKim2008}. The highly anisotropic crystalline structure leads to strong magnetic anisotropy(Fig.\ref{CCD}(a)). The system starts to develop three-dimensional anteferromagnetic(AFM) ordering below $T_N\sim 240~K $\cite{Crawford1994}, from the cooperation of a strong intraplane exchange $J$ of approximately 60 meV and a weak inter-plane exchange $J_c$ of approximately $16~\mu$eV \cite{JKim2012, Porras2019}. The magnetic ordering in \SIO{} has been well observed with x-ray resonant magnetic scattering (XRMS) measurements, appearing as Bragg-peaks in the magnetic scattering channel when the incident x-ray energy is tuned to $\sim$11.216 keV at the Ir-L$_{3}$ resonance edge\cite{BKim2009}. This high x-ray resonant energy allows the access to a large reciprocal space, thus the flexibility in selecting low x-ray incident angle scattering geometry to best meet the x-ray and laser penetration match requirement\cite{Guzelturk2021, Mark2016}. Our experiment was set up such that our $\sim$100 nm epitaxial thin film sample was stimulated with 1 eV laser ($\sim$100 fs) pulse by pulse, and the AFM ordering peak was continuously monitored with XRMS at an x-ray incident angle of less than $5^{\circ}$ to track the recovery of the magnetic ordering at a readout frequency of 1 Hz (see Ref. \cite{suppl}). 

\section{Experimental observations}

Figure \ref{CCD}(b) shows the (1 0 16) 3D AFM ordering peak in \SIO{} observed on the pixelated area detector at 80 K\cite{Lupascu2014}, and Fig.\ref{CCD}(c) depicts the detector surface projection in the reciprocal space. In order to catch the true first laser-shot response from pristine thermal-equilibrium condition, each dataset is collected after the sample goes through a fresh thermal cycle by warming above $T_N$ and then cooling back again to 80 K with the laser turned off. Then x-ray is turned on to continuously monitor the magnetic Bragg-peak intensity. In between, laser stimuli are controlled to come in pulse by pulse. The arrival time of the first laser pulse is defined as time zero. The evolution of the experimentally observed magnetic-peak height as a function of the laser fluence is shown in Fig.\ref{CCD}(d).   

After a single-laser-pulse stimulus to the magnetic ordering prepared from thermal-equilibrium evolution, the system does not recover to the initial state, evidenced by the permanent suppression of the magnetic Bragg-peak height as shown in Fig.\ref{CCD}(d). The degree of the suppression depends on the laser fluence, {\it i.e.}, the degree of the damage to the initial ordered spin network. More details of the response to the pulsed laser stimulation are shown in Fig.\ref{CCD}(e) with a laser fluence of 8.4 mJ/cm$^2$. At this fluence, the first laser pulse leads to an approximately $33\%$ drop in the magnetic Bragg-peak height, and the system settles at this condition without further evolution within an hour of continuous x-ray measurement. A second pulse arriving after one hour causes further suppression to the measured peak height, but with a much smaller drop. After 5$\sim$6 pulses, the system recovery becomes repeatable and more stimulation does not cause further permanent peak-height reduction. 

\begin{figure*}
\includegraphics[width=0.9\textwidth]{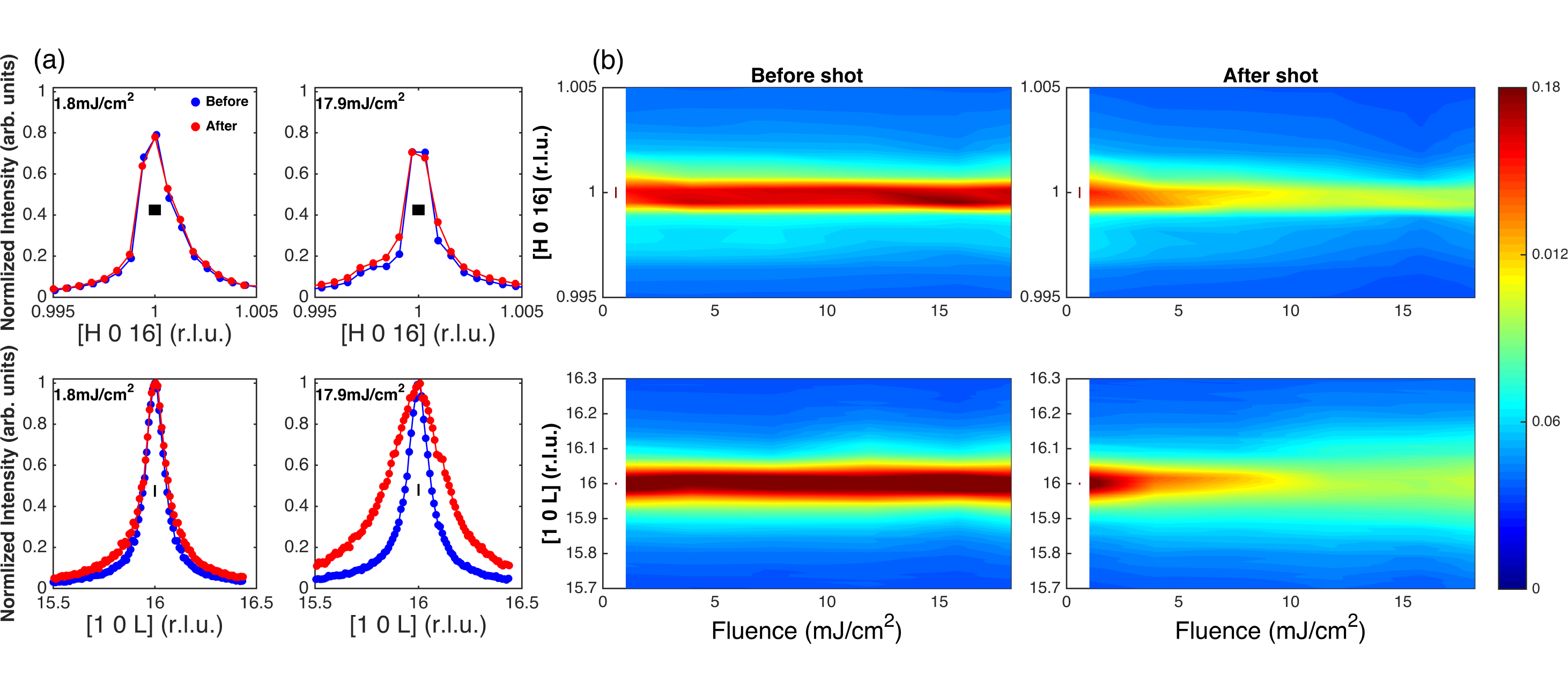}
\captionsetup{justification=raggedright,singlelinecheck=false}
\caption{\label{fig:wide}Evolution of the magnetic ordering upon laser stimulation. (a) Comparison of the $H$ (in-plane) and $L$ (interplane) scans before and after single-pulse-laser pumping. The horizontal bars are our instrumental resolution. Scans are normalized to the peak height determined from $L$ scans to emphasize the variation in the peak widths. (b) Fluence-dependence evolution of the $H$- and $L$- scans.}
\label{fig:HLscans}
\end{figure*}

These experimental observations clearly demonstrate that the recovery of the magnetic ordering in {\SIO} depends on the history of laser stimulation, and the response can be classified into two distinct stages. In the first stage of the initial few pulses, the degree of magnetic ordering keeps to be permanently suppressed and the system does not recover to the condition before pulse arrival. In the second stage, the system does recover but only to a state prepared by multiple initial laser pulses, which, of course is different from the thermal-equilibrium evolution. The laser-pulse-induced suppression of the magnetic ordering can be erased with a full thermal cycle and our measurements are fully reproducible (see Ref. \cite{suppl}), indicating the observed response is not due to sample damage but intrinsic to the spin system in \SIO. Phenomenologically, the observed shot-by-shot dependence is similar to the photoinduced metastable insulator-to-metal phase transition in La$_{2/3}$Ca$_{1/3}$MnO$_3$\cite{JZhang2016}. The laser stimulation drives the system into a nonthermal-equilibrium condition, and the deviation depends on the history of laser stimulation, a non-Markovian-type behavior. Such deviation is obviously beyond the classic three-temperature model\cite{Beaurepaire1996, QZhang2006,Koopmans2010, Kimling2014}.

To explore the microscopic origin of such evolution, detailed reciprocal space scans along the in-plane (H scans) and interplane (L scans) directions are performed across the magnetic peak before and after the first laser-pulse arrival under various laser fluences, and the results are shown in Fig.\ref{fig:HLscans}. Note that the peaks in the $H$ scan, a measure of the in-plane spin ordering, are very sharp and identical before and after the laser pulses for all laser fluences applied. From the width of the peaks in $H$ scans, the in-plane spin-ordering correlation length is estimated to be of a macroscopic scale about 0.9 $\mu m$. Thus, within a layer, the magnetic ordering is always fully restored to the thermal-equilibrium condition. The observed permanent suppression in the first stage is due to the incomplete recovery of the spin correlation along the interplane direction, which is evidenced by the broadening of the peaks in $L$ scans. With the spin-correlation length reduced, the scattering intensity is transferred from the center to the tails, leading to a permanent suppression of the peak height as observed. These results suggest that the laser pulse leads to highly anisotropic response between the in-plane and interplane AFM ordering correlations in the recovery process. 

\begin{figure*}
\includegraphics[width=0.85\textwidth]{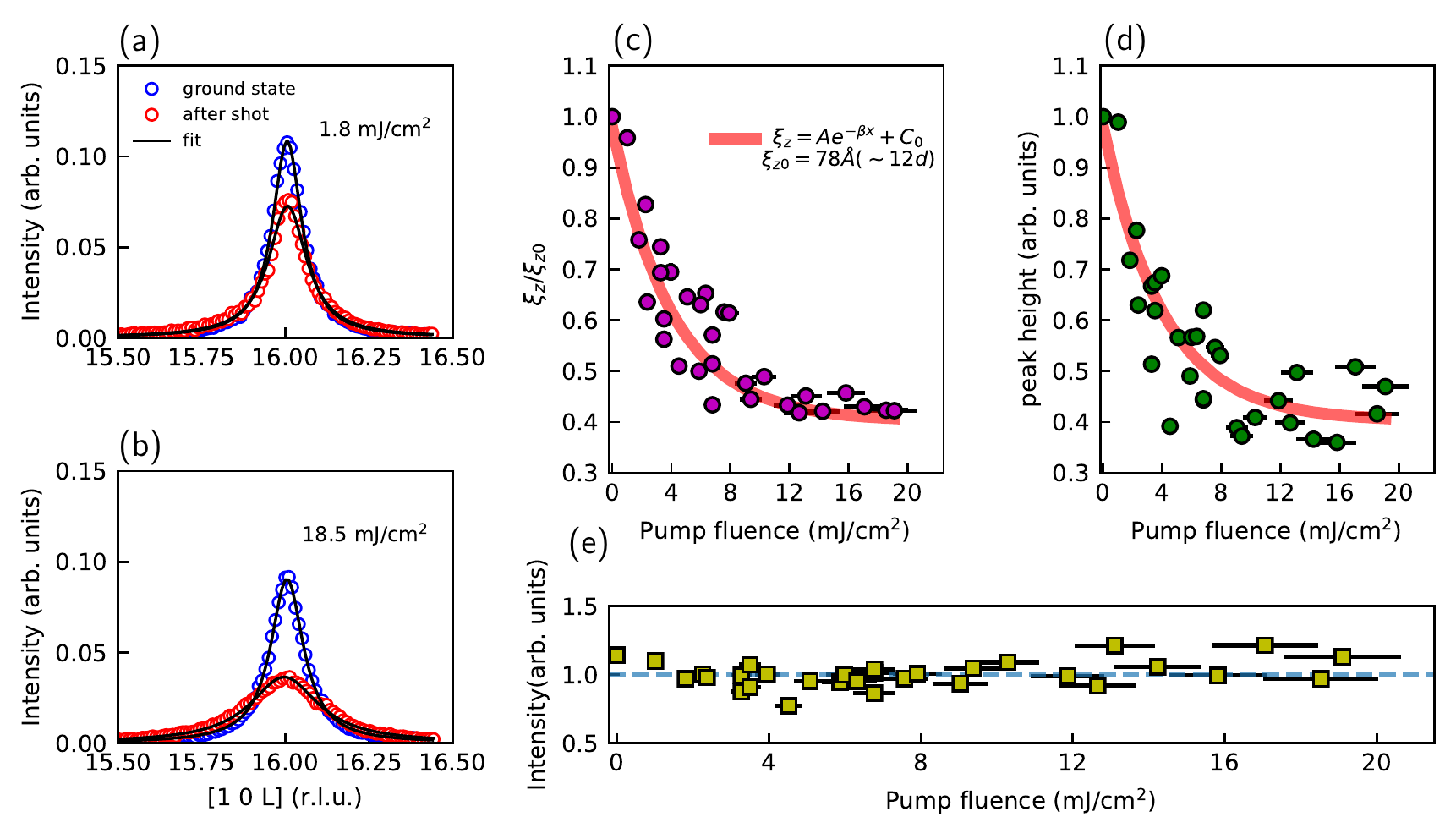}
\captionsetup{justification=raggedright,singlelinecheck=false}
\caption{\label{fig:wide}Evolution of the interplane correlation as function of fluence. (a,b) The fitting results for two representative conditions with weak and strong laser fluence. (c-e) The evolution of the interplane correlation length $\xi_z$, peak height and the integrated intensity as function of fluence, which are all normalized to the values before laser stimulation. With fluence higher than $\sim$12 mJ/cm$^2$, $\xi_z$ saturates to $\sim$5 layer thickness. The exponential lines are guides to the eye. The error bar for fluence is estimated to be $10\%$. (see Supplementary Material\cite{suppl})
}
\label{fig:fitting3}
\end{figure*}

To quantify such a response, the $L$ scans of the magnetic Bragg peak measured at thermal-equilibrium condition and after the very first laser pulse stimulus with varying fluence were analyzed to extract the evolution of the interlayer spin-ordering texture. It turns out that, for all experimental conditions, the $L$ scans can be well described by a single Lorentzian function as $I(q_z,\xi_z) \sim \frac{2}{d}\frac{\xi_z}{1+q_z^2\xi_z^2}$ (Fig.\ref{fig:fitting3}(a,b), and Ref. \cite{suppl}), with $d$ to be the interlayer distance and $q_z$ the interlayer direction relative-momentum transfer. The evolution of the extracted correlation length $\xi_z$ is shown in Fig.\ref{fig:fitting3}(c). As a function of the laser fluence, $\xi_z$, as well as the peak height (Fig.\ref{fig:fitting3}(d)), generally follows an exponential decay. On the other hand, the total area under the magnetic peak is conserved (Fig.3(e)). The significant data fluctuation mostly comes from slight misalignment in the H direction where the peak width corresponds to 0.01$^{\circ}$ in instrument rotational angle. Such conservation indicates that the size and direction of the local ordered magnetic moments are the same as those of the pristine condition\cite{Collins1989, Shannon1992}. Thus the incomplete recovery is solely due to the reduction in the interplane correlation. It is interesting to notice that the exponential decay is offset from zero. Both the correlation length and the peak height show a tendency to saturation at higher laser fluence, and the interplane spin correlation cannot be completely destroyed.

\section{Discussion}

Our observations reveal a few interesting aspects of the recovery of the magnetic-ordering texture in this highly magnetically anisotropic layered {\SIO}. First, starting from the thermal-equilibrium condition, each of the first few femtosecond-laser pulses induces permanent suppression to the magnetic ordering, and the degree of the suppression depends on the states prepared by the preceding pulses. Second, the system does eventually stabilize into conditions that the recovery of the magnetism becomes repeatable upon consecutive laser stimulation. This justifies the validity of experiments of stroboscopic mode in probing the spin dynamics\cite{RMP2010, Beaurepaire1996, Mark2016}. Third, although the degree of the partial recovery keeps degrading with increased laser fluence, it saturates to a robust nonzero value, indicating a minimum interplane partial recovery is intrinsically protected. The observed history and laser fluence- dependence hint at the nature of the magnetic recovery in {\SIO} upon ultrafast laser stimulation. 

The history dependence evidences that the demagnetization has certain local character. In the ultrafast process, the demagnetization does not completely wipe out all traces of magnetic order, and in the recovery the spins do not have long enough time to reach a global bath temperature. As a result, there are unperturbed (or less-perturbed) local spin clusters which preserve the memory of the prior state. Thus the prevailing global energy-dissipation picture in the classic three-temperature model\cite{Beaurepaire1996, QZhang2006, Koopmans2010,  Kimling2014} is oversimplified.

Notably, in both the initial and the second stages, the in-plane spin correlation always recovers to the thermal-equilibrium evolution condition of micron size. This differentiates our observation from the conventional fast quenching of a high-temperature state where the correlations along all directions are expected to change\cite{Bray1994, Mazenko1990, Shah1990}. The full in-plane recovery and the permanent loss of the interplane spin correlation are intimately related to the individual terms in the exchange interaction governing the magnetic dynamics along different directions. Although the 3D magnetic ordering in {\SIO} is jointly determined by both the intraplane and interplane magnetic couplings \cite{Kastner1998,Porras2019,Fujiyama2012}, in the ultrafast recovery process they act differently. The strong in-plane exchange of tens of meV drives a quick re-establishment of the in-plane correlation within a few picoseconds\cite{Mark2016}. We suggest that in a such short duration, the additional energy introduced by laser pumping into the spin system cannot be efficiently absorbed by the lattice reservoir\cite{JLi2019, Yuelin2016}. Instead, the spin sector is still highly excited in picosecond time scale. Associated with the weak interlayer exchange $J_c$ of approximately $16~\mu$eV, a time window much longer than picoseconds is needed to allow them to fully dissipate. Such a process is cut off when the macroscopic intralayer spin correlation is established, since there is an enormous energy barrier to flip a whole layer.  

\begin{figure}[h]
\centering
\includegraphics[width=0.45\textwidth]{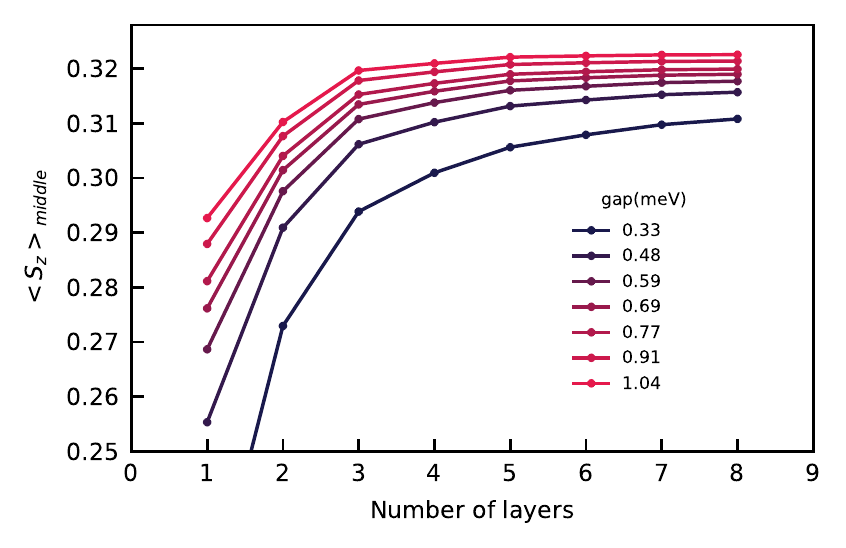}
\captionsetup{justification=raggedright,singlelinecheck=false}
\caption{Theoretical modeling. Calculated magnetic-order parameter at the middle of the slab, $\langle S_{z}\rangle_{middle}$, as a function of the layer number from the model Hamiltonian $H = \frac{J}{2}\sum_{l,ij}\Vec{S}_{li}\cdot\Vec{S}_{lj}+\Delta\sum_{l,ij}S^{z}_{li}S^{z}_{lj} +\frac{J_c}{2}\sum_{ll^{'},ij}\Vec{S}_{li}\cdot\Vec{S}_{l^{'}j}$ at T = 80 K with exchange terms $J= 60~meV$, $Jc = 16.4~ \mu eV$. Different curves correspond to different anisotropic exchange $\Delta$ strengths, and the consequent gap sizes are used as an index.}
\label{order_calculation}
\end{figure}

Following the above arguments, we can understand the saturation plateau at high laser fluence as shown in Fig.\ref{fig:fitting3}(c,d). In our earlier report\cite{Mark2016}, we have demonstrated that the laser pulse completely destroys the magnetic order in the initial hundreds of femtoseconds, when the laser fluence is stronger than approximately 12 mJ/cm$^2$. Thus, beyond such laser fluence, the system completely loses its memory of the history and the recovery follows the same path, regardless of further laser-fluence increasing.
This is also consistent with the observed multishot evolution where, at high fluence, the laser pulses after the very first shot drive marginal further suppression to the magnetic Bragg-peak height (see Fig. S9 in Supplemental Material \cite{suppl}). Without the assistance from the remnant order, reestablishment of the global interplane correlation completely lags behind the intraplane recovery. Thus the system recovery enters a quasi-two-dimensional regime. For such condition, Mermin and Wagner\cite{Mermin1966} prove that spontaneous two-dimensional ferromagnetic or antiferromagnetic long-range order at finite temperature is highly susceptible to spin thermal fluctuations, and the nonzero third-dimension correlation is critical to suppress the thermal fluctuation to realize long-range two-dimensional ordering. 

We confirm such an explanation for our observed plateau by calculating the magnetic correlation function in a few layers of square lattice 
from an effective anisotropic Heisenberg spin-$1/2$ model, whose dynamics is solved with the equation-of-motion technique and mean-field approximation\cite{Diep1991}(see Ref. \cite{suppl}). With the realistic parameters from experiment and published literature\cite{JKim2012,Porras2019}, the self-consistent ordered magnetic moment $\langle S_{z}\rangle$ and the magnetic correlation function $\langle S^{-}S^{+}\rangle$ are obtained as functions of the model slab thickness. As the reported exchange anisotropy for \SIO{} is quite controversial\cite{Porras2019, YGim2016, Calder2018, Pincini2017}, the calculated $\langle S_{z}\rangle_{middle}$, the ordered magnetic moment at the middle of the slab, is shown in Fig.\ref{order_calculation} as a function of the calculated magnon gap size. As expected, both the anisotropic exchange $\Delta$ strengths and the layer thickness are critical for true long-range magnetic ordering. When the gap size approaches approximately 1 meV, the magnetic-order parameter $\langle S_{z}\rangle_{middle}$ stabilizes around 4-5 layers, which agrees with the interplane correlation length of the saturation plateau we observe in experiment (Fig.\ref{fig:fitting3}(c,d)). Thus, we realize a spin thermal fluctuation limit in real material with laser-pulse stimulation. Furthermore, this result indicates that indeed the interplane recovery-time window is set by the in-plane spin correlation. The longer interplane correlation established at lower fluence is assisted by memory of the spin network, which is only partially destroyed below the high-fluence threshold.

\section{Conclusion}

In conclusion, a history and laser-fluence dependence of the partial-recovery process of the 3D AFM ordering in \SIO{} was observed, which is related to the distinctly different timescales for the interplane and intraplane recoveries in the nonthermal-equilibrium ultrafast process. The noncooperation of the different exchange interactions in the ultrafast process drives the system to deviate from the quasiequilibrium relaxation pathway. Light-induced deviation from thermal-equilibrium evolution has been rarely observed, and previous reports are associated with the lattice\cite{Stoica2019} or charge\cite{Kiryukhin1997, JZhang2016, Stojchevska2014,SuYang2020} degrees of freedom. Our results extend the direct observation of deviation from recovery
to thermal-equilibrium condition into the spin sector. Furthermore, we suggest that such time-window mismatch could generally happen in complex systems during ultrafast nonthermal evolution, and our observations could be relevant to a wide range of problems in the nonequilibrium-behavior of low-dimensional magnets and related ordering phenomena. For example, the laser-induced `hidden quantum state' in layered {\it 1T}-TaS$_2$ could be one special case\cite{Stojchevska2014, SLee2019}.

\section{acknowledgments}
We thank Yi Zhu for the assistance of experimental setup. The experimental work by X. L. and R. W. was primarily supported by National Natural Science Foundation of China under grant No. 11934017. Part of the execution of the experiment and data interpretation by H. W. and Y. C. were supported by the U.S. DOE, Office of Science, Office of Basic Energy Sciences, Materials Sciences and Engineering Division. Use of the Advanced Photon Source was supported by the U.S. Department of Energy, Office of Science, Office of Basic Energy Sciences under contract No. DE-AC02-06CH11357. Work at Brookhaven National Laboratory was supported by the U.S. DOE, Office of Science, Office of Basic Energy Sciences, Materials Sciences and Engineering Division under Contract No. DE-SC0012704. J. L. acknowledges support from the National Science Foundation under Grant No. DMR-1848269. H. D. acknowledges support from the National Natural Science Foundation of China (No. 11888101). H. D. and X. L. acknowledge support from the Ministry of Science and Technology of China (2016YFA0401000).  J. Y. acknowledges funding from the State of Tennessee and Tennessee Higher Education Commission (THEC) through their support of the Center for Materials Processing.

\end{document}




\title{Supplemental material for ``Single Laser Pulse Driven Thermal Limit of the Quasi-Two Dimensional Magnetic Ordering in \SIO"}

\author{Ruitang Wang$^{1,2,3}$}
\author{J. Sun$^{1}$}
\author{D. Meyers$^{4,5}$}
\author{J. Q. Lin$^{1,2,3}$}
\author{J. Yang$^{6}$}
\author{G. Li$^{1}$}
\author{H. Ding$^{2,3}$}
\author{Anthony D. DiChiara$^{7}$}
\author{Y. Cao$^{8}$}
\author{J. Liu$^{6}$}
\author{M. P. M. Dean$^{4}$}
\author{Haidan Wen$^{7,8}$}
\author{X. Liu$^{1}$} 
 \email{liuxr@shanghaitech.edu.cn}
\affiliation{$^{1}$School of Physical Science and Technology, ShanghaiTech University, Shanghai 201210, China.}
\affiliation{$^{2}$
Beijing National Laboratory for Condensed Matter Physics and Institute of Physics, Chinese Academy of Sciences, Beijing 100190, China}
\affiliation{$^{3}$
University of Chinese Academy of Sciences, Beijing 100049, China}
\affiliation{$^{4}$
Condensed Matter Physics and Materials Science Department, Brookhaven National Laboratory, Upton, New York 11973, USA.}
\affiliation{$^{5}$
Department of Physics, Oklahoma State University, Stillwater, Oklahoma 74078, USA.}
\affiliation{$^{6}$
Department of Physics and Astronomy, University of Tennessee, Knoxville, Tennessee 37996, USA.}
\affiliation{$^{7}$
Advanced Photon Source, Argonne National Laboratory, Argonne, IL, 60439, USA.}
\affiliation{$^{8}$
Materials Science Division, Argonne National Laboratory, Argonne, Illinois, 60439, USA.}

\pacs{Valid PACS appear here}
\maketitle


\maketitle

\section{\label{sec:level2}Sample Synthesis and Characterization}

\begin{figure}[b] 
\renewcommand{\figurename}{Fig. S}
\centering
\includegraphics[width=0.45\textwidth]{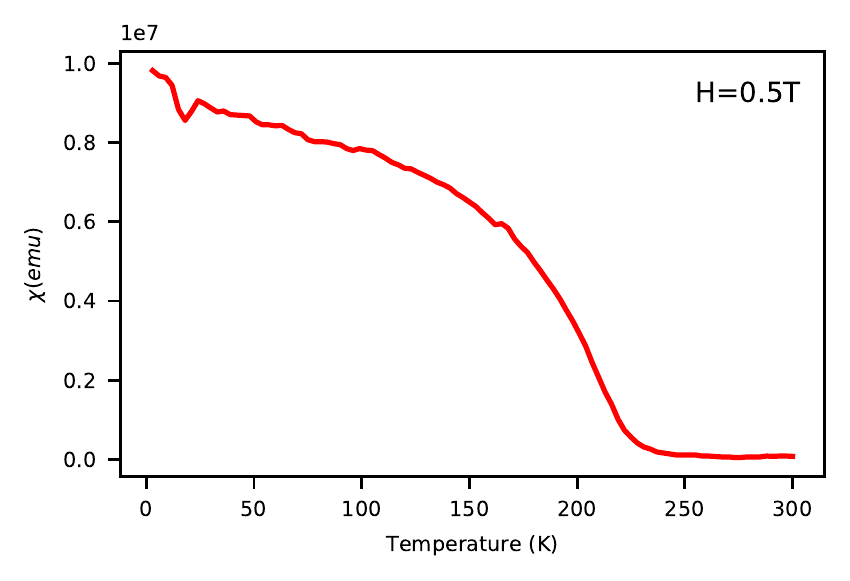} 
\captionsetup{justification=raggedright,singlelinecheck=false}
\caption{Temperature dependence of the magnetic susceptibility of our Sr2IrO4 thin film sample.}
\label{fig:magnetic susceptibility}
\end{figure}

\begin{figure*}[htb]
\renewcommand{\figurename}{Fig. S}
\centering
\includegraphics[width=0.9\textwidth]{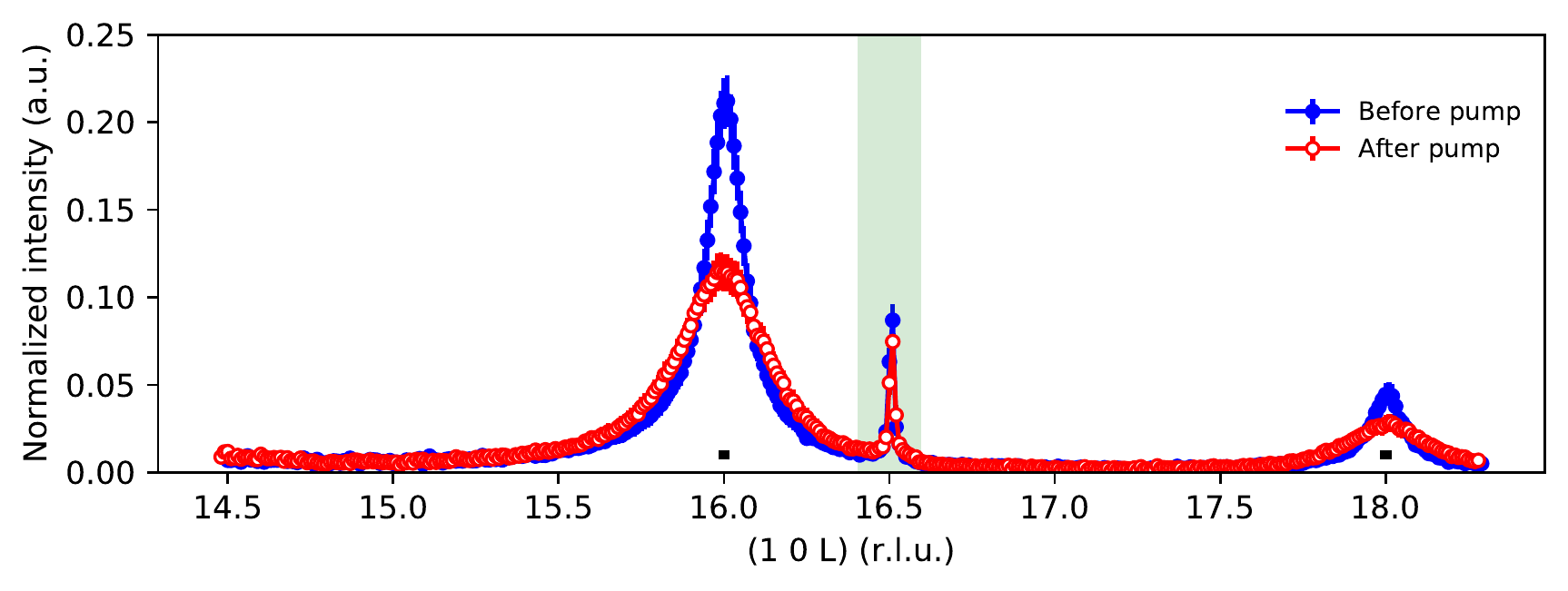}
\captionsetup{justification=raggedright,singlelinecheck=false}
\caption{Long-range L-scan of (1 0 L) magnetic Bragg peaks along c-axis direction. The long-range reciprocal space scan along out-of-plane direction(L scan) was performed for pristine thermal equilibrium condition of the sample at 80K(blue) and after the first single laser shot(red). The appearance of both (1 0 16) and (1 0 18) magnetic Bragg peaks indicate the existence of twinned magnetic domains([(1 0 4n),(0 1 4n+2)] and [(1 0 4n+2),(0 1 4n)]) in our sample. Both of these two magnetic domains respond to the single shot in a similar manner. The horizontal small bars represents the intrumental resolution. The peak at $L=16.5$(marked by the shaded region) comes from the ($\frac{1}{2}$ $\frac{1}{2}$ $\frac{5}{2}$) superlattice peak of the SrTiO$_{3}$ substrate.}
\label{fig:longLscan}
\end{figure*}

Sr$_2$IrO$_4$ thin film samples with thickness of $\sim$100 $nm$ grown with PLD method were used\cite{Rayan2013}. \SIO\ is crystallized in $I4_{1}/acd$ structure with single IrO$_2$ layers separated by SrO layers\cite{Huang1994}. Each structural unit cell contains two Ir in one layer and four IrO$_2$ layers along c direction. The antiferromagnetic(AFM) ordering sets in at $T_{N}\approx 240K$ for bulk crystal. In our thin film samples, the magnetic susceptibility measurement gave slightly lower $T_{N}$ as shown in \textbf{Fig. S\ref{fig:magnetic susceptibility}}. The AFM ordering shares the same unit cell at that of the structure\cite{Fujiyama2012}. In the tetragonal \SIO, two twined magnetic domains are expected to produce two sets of magnetic reflection peaks at (1 0 4n), (0 1 4n+2) and (1 0 4n+2), (0 1 4n) respectively, where lattice reflections are forbidden. The magnetic ordering peaks from both domains were observed in the long range L-scan shown in \textbf{Fig. S\ref{fig:longLscan}}, which are at (1 0 16) and (1 0 18) respectively. Importantly, both peaks respond to laser stimulation in the same way. 

\section{\label{sec:citeref}Characterization of Laser Spot and Estimation of the Effective Average Fluence}
Discrete laser pulses with a duration of $\sim 100$ fs was used to pump the sample. The pump laser photon energy was selected to be 1 eV derived from a Ti: Sapphire laser system with an optical parametric amplifier, corresponding to the resonant excitation from $J_{eff}=\frac{3}{2}$ to unoccupied $J_{eff}=\frac{1}{2}$ states \cite{Moon2009}. Thus the pumping largely creates double occupancy of the $J_{eff} = \frac{1}{2}$ states and leaves a hole in the $J_{eff} = \frac{3}{2}$ manifold. 
We have shown that 1eV pumping can efficiently break the long range AFM ordering\cite{liu-nmat2016}.
The laser system runs at 1 KHz and can be controlled to deliver a single-shot laser pulse on demand.

\begin{figure}[b]
\renewcommand{\figurename}{Fig. S}
\centering
\includegraphics[width=0.45\textwidth]{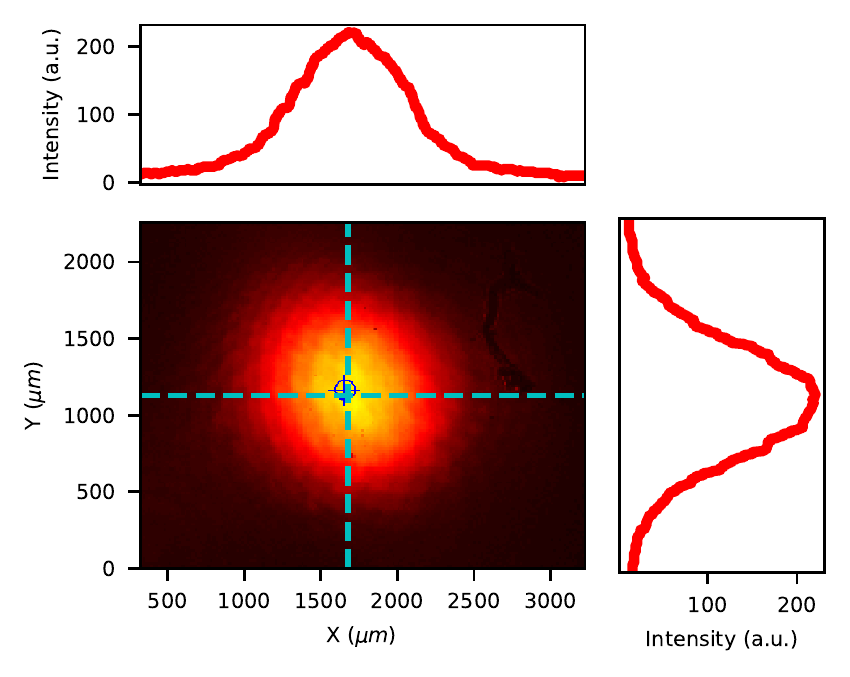}
\captionsetup{justification=raggedright,singlelinecheck=false}
\caption{Characterization of the laser spot size. The energy density profile of the pump laser was measured with CCD, as shown in the image. Two line cuts (shown by the dashed lines in the CCD image) were taken to extract the peak widths. From Gaussian fitting, both directions give $\sigma = 350\mu m$.}
\label{fig:laserSpot}
\end{figure}

To properly characterize the laser fluence, the laser power density profile was measured, as shown in \textbf{Fig. S\ref{fig:laserSpot}}. The profile can be be modeled as an isotropic Gaussian pulse with the fitted $\sigma$ to be 350$\mu$m. In the experiment, the laser incident angle was $47^{\circ}$. Thus the laser on-sample footprint was elongated along one direction with $\sigma' = \sigma/sin(47^{\circ})=479\mu$m. The effective laser fluence under the X-ray spot can be calculated as,    

\begin{equation}
F = \frac{P}{Af}\iint_{A}\frac{1}{2\pi\sigma\sigma'}exp[-(\frac{x^2}{\sigma^2} + \frac{y^2}{\sigma'^2})]dxdy
\label{eqn:Fluence}
\end{equation} 
where A is the overlapped area of the X-ray at the laser spot center on the sample surface, $f$ is the running frequency as 1 kHz, and $P$ is the laser power measured during the experiments. At $P = 1 mW$ with $A = 2.41\times10^{-4}  cm^2$, the average fluence within the overlapped region of the single laser shot and X-ray beam spot on the sample is 0.113 $m J/cm^2$. As the X-ray to the laser spot center overlap was done by referring to a video camera monitor, we expect certain miss-alignment. In \textbf{Fig. 3} in the main text, the error bar given for fluence is $10\%$ by assuming possible $\pm 60\mu m$ miss-alignment. 

At $\sim 1$eV, the penetration depth of the pumping laser for \SIO{} is estimated{\cite {liu-nmat2016}} to be $\sim$100 nm.

\section{Schematic of Experiment}%
\begin{figure}[htb]
\renewcommand{\figurename}{Fig. S}
\centering
\includegraphics[width=0.45\textwidth]{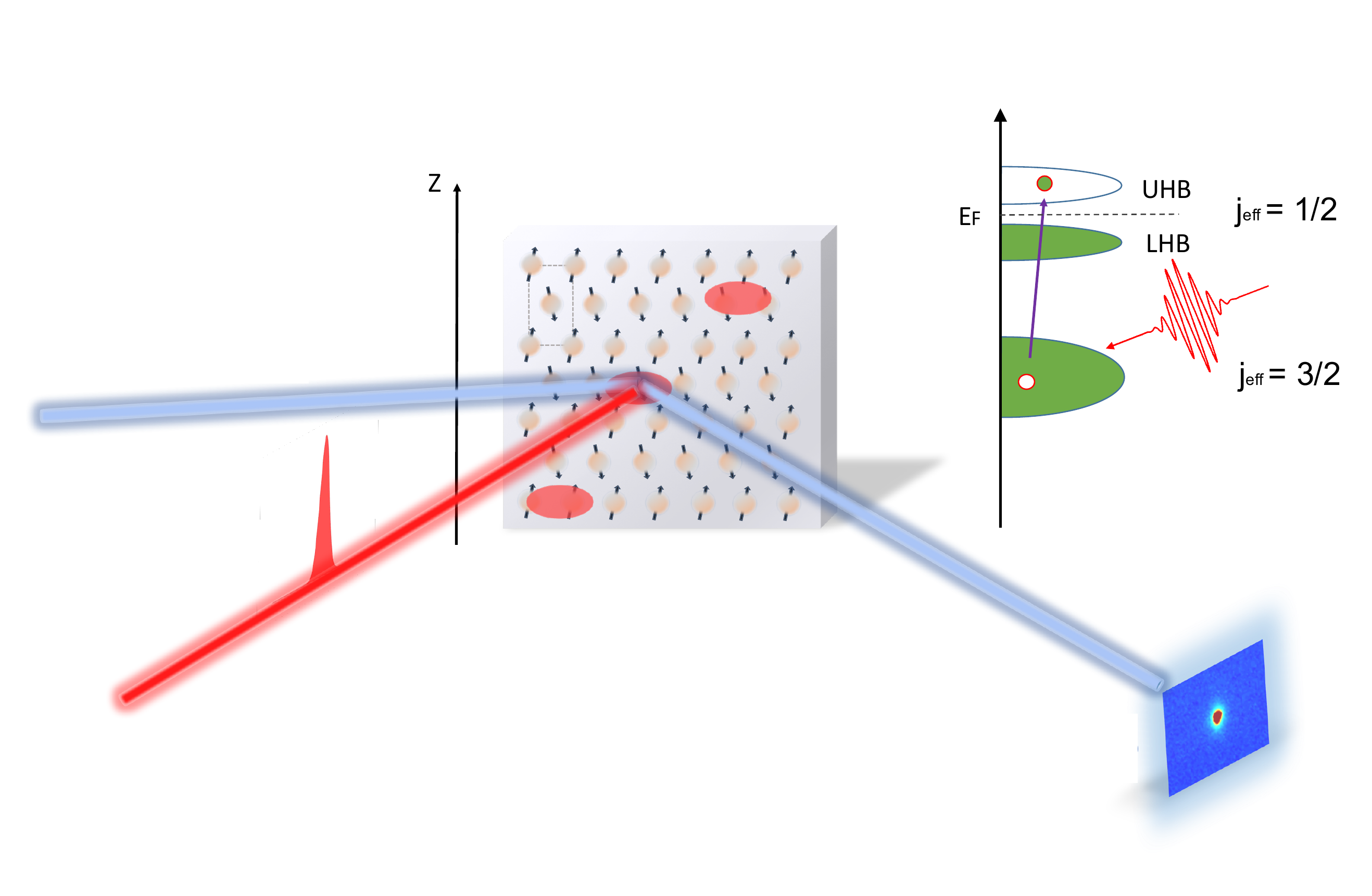}
\captionsetup{justification=raggedright,singlelinecheck=false}
\caption{Experimental configuration. The XRMS experiment was performed in a horizontal scattering geometry where the a- and c-axis of sample lie within the scattering plane. Three far-apart separated spots on the sample were chosen to be measured in between slow thermal-cycle processes to save experimental time. The inset shows the corresponding electron excitations from laser pumping, where 1 eV laser pulses mainly drive electrons from $J_{eff}=\frac{3}{2}$ states into the unoccupied $J_{eff}=\frac{1}{2}$ states. }
\label{fig:geometry}
\end{figure}

The X-ray resonant magnetic scattering(XRMS) measurements were conducted at the Advanced Photon Sources(APS) using beamline 7-ID-C. The data were collected at Ir $L_3$ absorption edge of 11.216 KeV. A horizontal scattering geometry was used.(see \textbf{Fig. S\ref{fig:geometry}}) The laser pulse came in at a large angle of $47^{\circ}$ relative to the sample surface to allow a more homogeneous excitation along the sample depth direction.  

\begin{figure}[htb]
\renewcommand{\figurename}{Fig. S}
\centering
\includegraphics[width=0.45\textwidth]{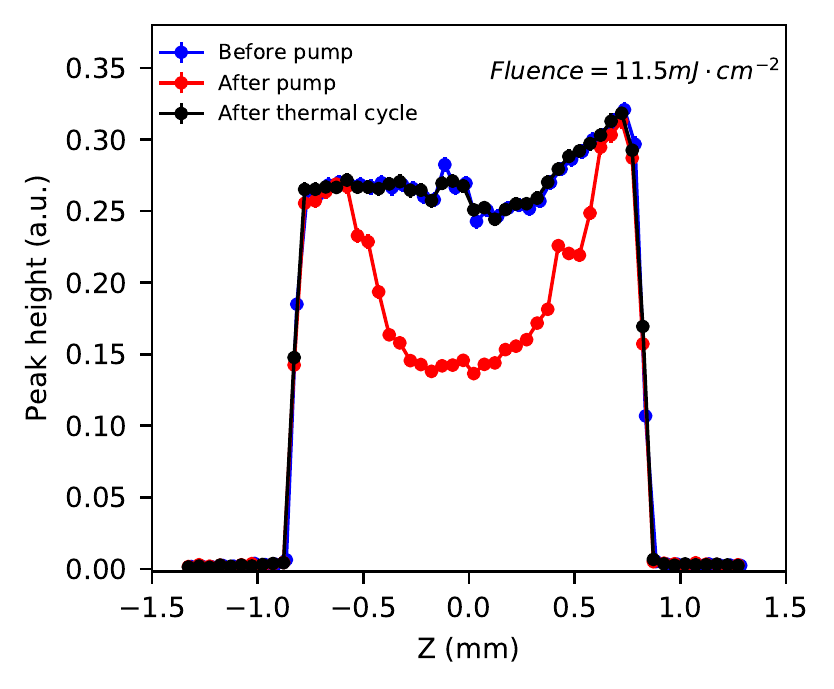}
\captionsetup{justification=raggedright,singlelinecheck=false}
\caption{Cross-sample scan. Real space scans along z-direction(see the experimental schematic in \textbf{Fig. S\ref{fig:geometry}}) across the sample while monitoring the (1 0 16) magnetic peak height. Three scans were taken for:  pristine thermal equilibrium condition(blue curve), after single laser pulse excitation(red curve), and after a full thermal cycle process(black curve). The laser pumping leads to a drastic suppression of the scattering intensity, which is fully recovered after a thermal cycle.}
\label{fig:thermalCycle}
\end{figure}

To amplify the magnetic scattering signal\cite{Liu_Na2IrO3}, the scattering experiment was performed in the $a$-$c$ plane with the incident X-ray came in at a shallow angle of $4.87^{\circ}$ relative to the sample surface. Its polarization was almost parallel to the sample surface $c$-direction. A Pilatus CCD with pixel size of 172 $\mu m^{2}$ was used in the experiment to monitor the 
scattered X-ray signal. It was placed $\sim 1m$ away from the sample, which gives an angular resolution of 0.01$^\circ$ per pixel. The sample was cooled down to 80 K with cryostat, well below the N\'eel ordering temperature. During the experiment, a full thermal cycle was done by warming the sample up to 280 K and then slowly cooling down to 80 K with a cooling rate of 0.05 K/s. The laser induced suppression of the magnetic peak height was fully recovered after a thermal cycle, as shown in the cross-sample scan in \textbf{Fig. S\ref{fig:thermalCycle}}. The entire process is repeatable, ruling out the irreversible sample damage issue.

\section{\label{sec:level2}Laser Effect on Structural Peak}
\begin{figure}[b] 
\renewcommand{\figurename}{Fig. S}
\centering
\includegraphics[width=0.45\textwidth]{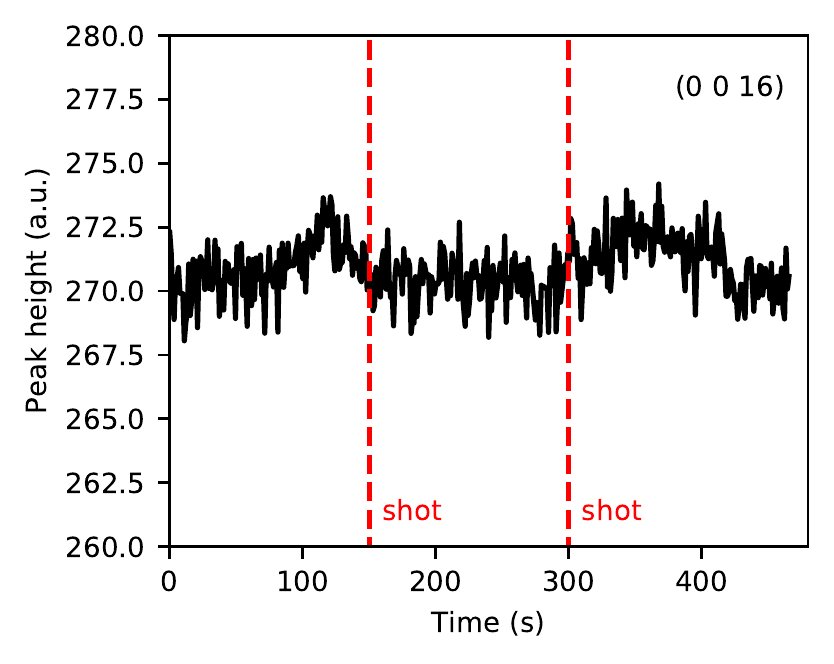} 
\captionsetup{justification=raggedright,singlelinecheck=false}
\caption{Laser effect on structural peak. response of the (0 0 16) structural Bragg peak height to laser stimulation. The dashed lines mark two single laser pulses.}
\label{fig:struc_peak}
\end{figure}

To check the laser shot effect on the crystal structure, (0 0 16) structure peak height was monitored with single laser pulse stimulation. As shown in \textbf{Fig. S\ref{fig:struc_peak}}, the fluctuation in the structural peak height, mainly due to X-ray beam instability, is uncorrelated with the laser stimulation. 
Thus laser induces minimum effect to the lattice at 1Hz frequency of which our data was taken, and the suppression of the magnetic peak height is intrinsic to the spin sector.

\section{\label{sec:level2}X-ray Effect on Magnetic Peak}

\begin{figure}[b] 
\renewcommand{\figurename}{Fig.S}
\centering
\includegraphics[width=0.45\textwidth]{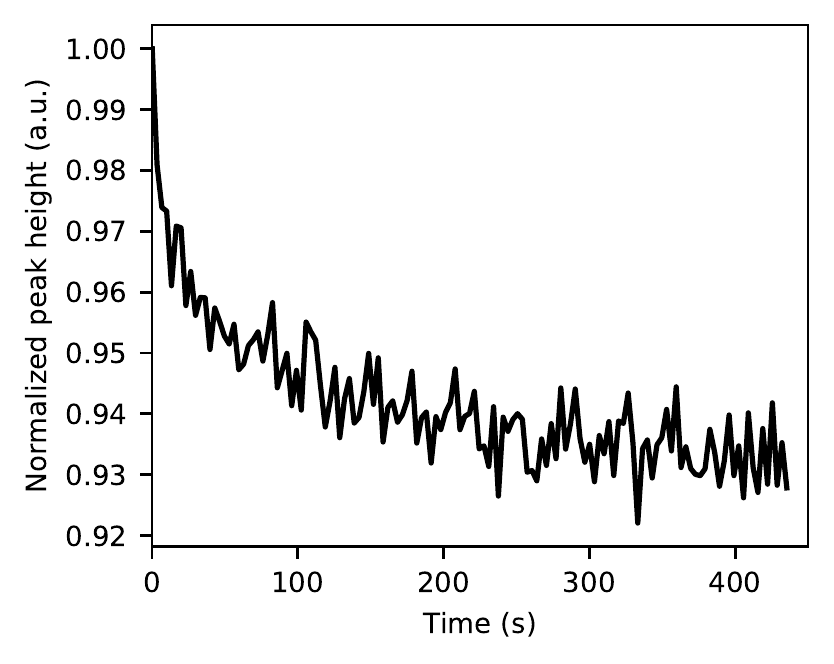} 
\captionsetup{justification=raggedright,singlelinecheck=false}
\caption{X-ray effect on magnetic peak. Temporal evolution of the (1 0 16) magnetic Bragg peak height (normalized) in the initial 420 seconds after turning on the X-ray.}
\label{fig:xray_effect}
\end{figure}

We checked the X-ray effect on magnetic peak by monitoring the (1 0 16) magnetic Bragg peak height after X-ray was initially turned on after a full thermal cycle, without any optical pumping on sample. A gradual reduction of the peak height about 7\% was noticed after the X-ray exposure of the sample, as shown in \textbf{Fig. S\ref{fig:xray_effect}}. Then the peak height stabilized after a few minutes.

\newpage
\section{Formula for the Diffraction Profile of the L-scan of Magnetic Bragg Peak}

Since the in-plane AFM ordering correlation length is fully restored, we focus on the the inter-plane correlation. A phenomenological model is constructed by assuming the AFM ordered iso-spins are still pointing to the crystal $a$-direction as in the thermal equilibrium state while their inter-plane correlation is described by an exponential decay as $e^{-\frac{|z_m-z_n|}{\xi}}$ with $z_{m(n)}$ to be the c-direction coordinate. Accordingly, the magnetic reflection intensity can be written as,

\begin{widetext}  \begin{eqnarray} 
I(Q_z,\xi_z) = \mid F \mid^2 \Big[\delta(Q_x-h\cdot a^*)  \delta(Q_y-k\cdot b^*)\Big]^2 \cdot \frac{1}{N_3}\sum_{j,k=1}^{N_3} (-1)^j e^{-iQ_zz_j}  (-1)^k e^{iQ_zz_k} e^{-\frac{|z_j-z_k|}{\xi_z}}
\end{eqnarray}  
\end{widetext}

where $F$ is the magnetic scattering factor for Ir sites, and $j$ and $k$ represents the $j$-th and $k$-th plane along c-direction. $N_3$ is the total plane number along c-direction. $\xi_z$ is the c-direction magnetic correlation length. The in-plane structure factors are simplified to $\delta$-functions due to the fact that the in-plane correlation lengths are orders of magnitude larger than the c-direction correlation length(see main text). The summation can be analytically carried out as:
\begin{equation}
I(Q_z,\xi_z)  = \mid F \mid^2\frac{sinh(\frac{d}{\xi_z})}{cosh(\frac{d}{\xi_z})-cos(Q_zd)}
\label{eqn:ana}
\end{equation} 
where $d$ is the inter-layer distance.   

The relative momentum transfer can be defined as $q_z=Q_z-G$ with G indexing the Bragg points.  When $\frac{(q_z d)^4}{4!} << 1$, close to the Bragg point as where our {\it L}-scans were taken,  
Eqn. \ref{eqn:ana} can be simplified as:

\begin{align}
I(q_z,\xi_z) &=\mid F \mid^2\frac{sinh(\frac{d}{\xi_z})}{cosh(\frac{d}{\xi_z})-cos(q_zd)} \\  
&\approx\mid F \mid^2\frac{(\frac{d}{\xi_z})}{\frac{d^2}{2\xi_z^2}+\frac{(q_z)^2}{2}}\\ 
&=\mid F \mid^2\frac{2}{d}\frac{\xi_z}{1+q_z^2 \xi_z^2}
\label{eqn:fit}
\end{align}

Thus, with inter-layer ordering correlation defined as $e^{-\frac{|z_m-z_n|}{\xi_z}}$, the X-ray scattering profile is of a Lorentzian shape. The peak height at $q_z = 0$ should be proportional to the correlation length $\xi_z$, while the whole integrated intensity is constant. All these predictions agree well with our observations, suggesting a quite homogeneous statistical distribution of the {\it c}-direction spin ordering.

\section{\label{sec:level2}Fitting Procedure of the Magnetic Bragg Peak}

\begin{figure*}[htb]
\renewcommand{\figurename}{Fig. S}
\centering
\includegraphics[width=0.9\textwidth]{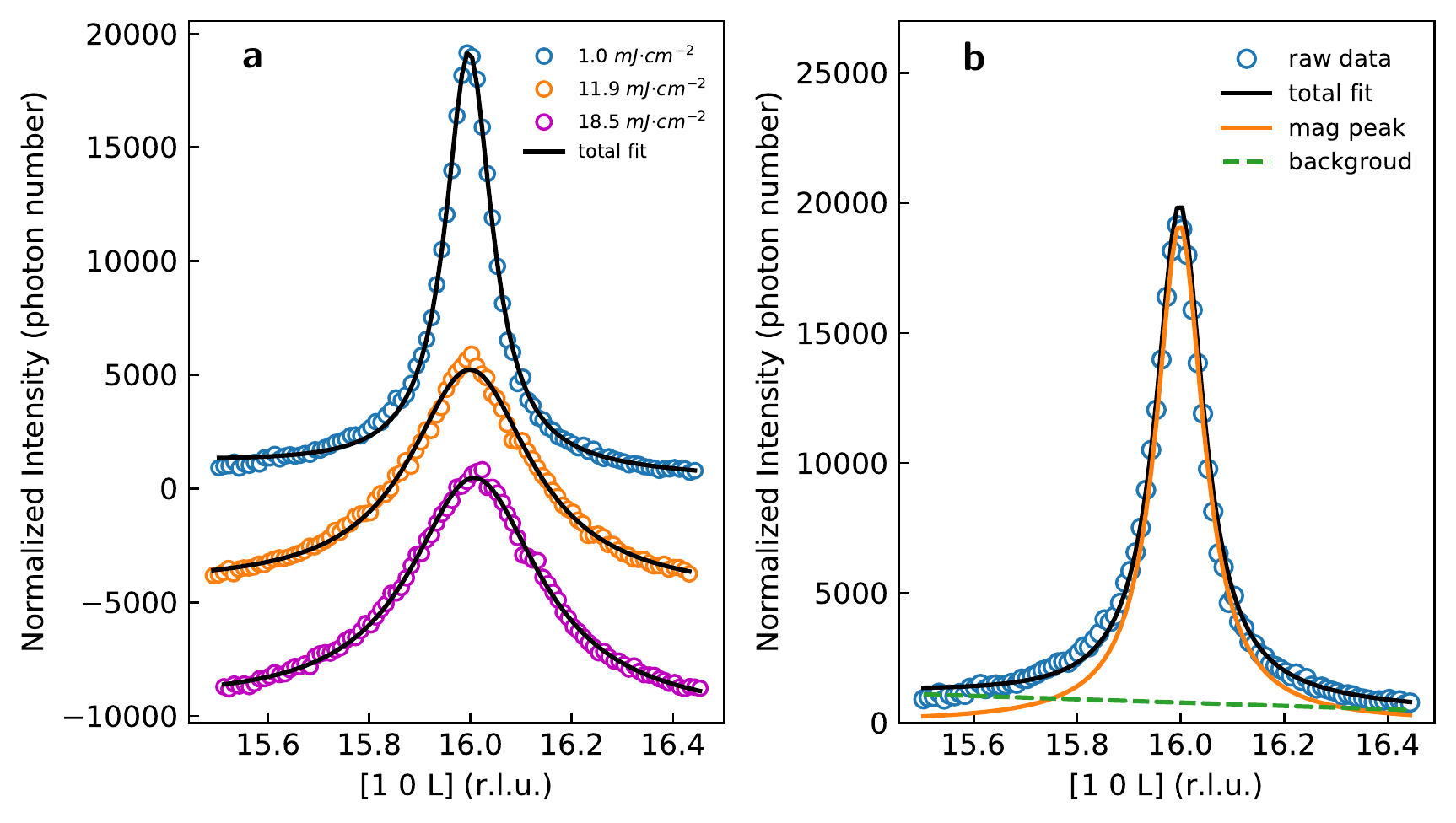}
\captionsetup{justification=raggedright,singlelinecheck=false}
\caption{\textbf{Fitting Procedure of Magnetic Bragg Peak:} All the magnetic Bragg peaks were fitted with the Lorentzian function we derived with a linear background(Eqn. \ref{eqn:fitt}) using the least-squares fitting method. \textbf{a}, Three L-scan data after laser excitation and the fitting results (vertically stacked for clarity). \textbf{b}, The fitting components of a typical L-scan data.}
\label{fig:fit_Lscan}
\end{figure*}

All the magnetic peaks were fitted based on the Eqn. \ref{eqn:fit} plus a linear background intensity, as shown in Eqn. \ref{eqn:fitt}. For a set of magnetic Bragg peaks studied with the same fluence of laser pulse, firstly we fit the magnetic Bragg peaks of pristine thermal limit(before laser excitation), and extract a background intensity; Then we fit the magnetic Bragg peaks after the single shot excitation with the same background intensity.(see \textbf{Fig. S\ref{fig:fit_Lscan}b}) 

\begin{equation}
I(q_z, h, \xi_z) = \frac{2}{d}\frac{h}{1+q_z^2 \xi_z^2} + I_{bg}
\label{eqn:fitt}
\end{equation}
The fitting was done by least-squares fitting. And here the reduced Chi-square $\chi_{\nu}^2$ is defined as:
\begin{equation}
    \chi_{\nu}^2 = \frac{1}{N-N_{varys}}\sum_{i}^N\frac{[y_{i}^{exp}-y_{i}^{model}(v)]^2}{\epsilon_{i}^2} 
\end{equation}

where $N$ is the number of data points, $N_{varys}$ is the number of variables in the fit, $y_{i}^{exp}$ is the measured data, $y_{i}^{model}(v)$ is the model calculation and $v$ is the set of variables in the model to be optimized in the fit, and $\epsilon_{i}$ is the estimated uncertainty in the data.

\begin{table*}
\centering
\caption{Fitting goodness}
\label{tab:fitresult}
\begin{tabular}{c rrrrrrrrr}
\hline\hline 
Fluence($mJ/cm^2$)\quad& 1.0  \quad&  3.3  \quad&  6.0  \quad&  6.8  \quad&  9.4 \quad& 11.9 \quad& 13.1 \quad& 18.5\\
\hline
$\chi_{\nu}^2$ \quad & 18.4 \quad& 21.6 \quad& 30.7 \quad& 8.5 \quad& 4.9 \quad& 8.7 \quad& 6.8 \quad& 6.2\\[1ex]

\hline
\end{tabular}
\end{table*}

Representative fitting goodness is listed in Table. S\ref{tab:fitresult}. And we plot three of the typical fitting results in \textbf{Fig. S\ref{fig:fit_Lscan}a}.

\section{\label{sec:level2}Multiple Shots Evolution of the Magnetic Peak Height}

\begin{figure*}[htb]
\includegraphics[width=0.9\textwidth]{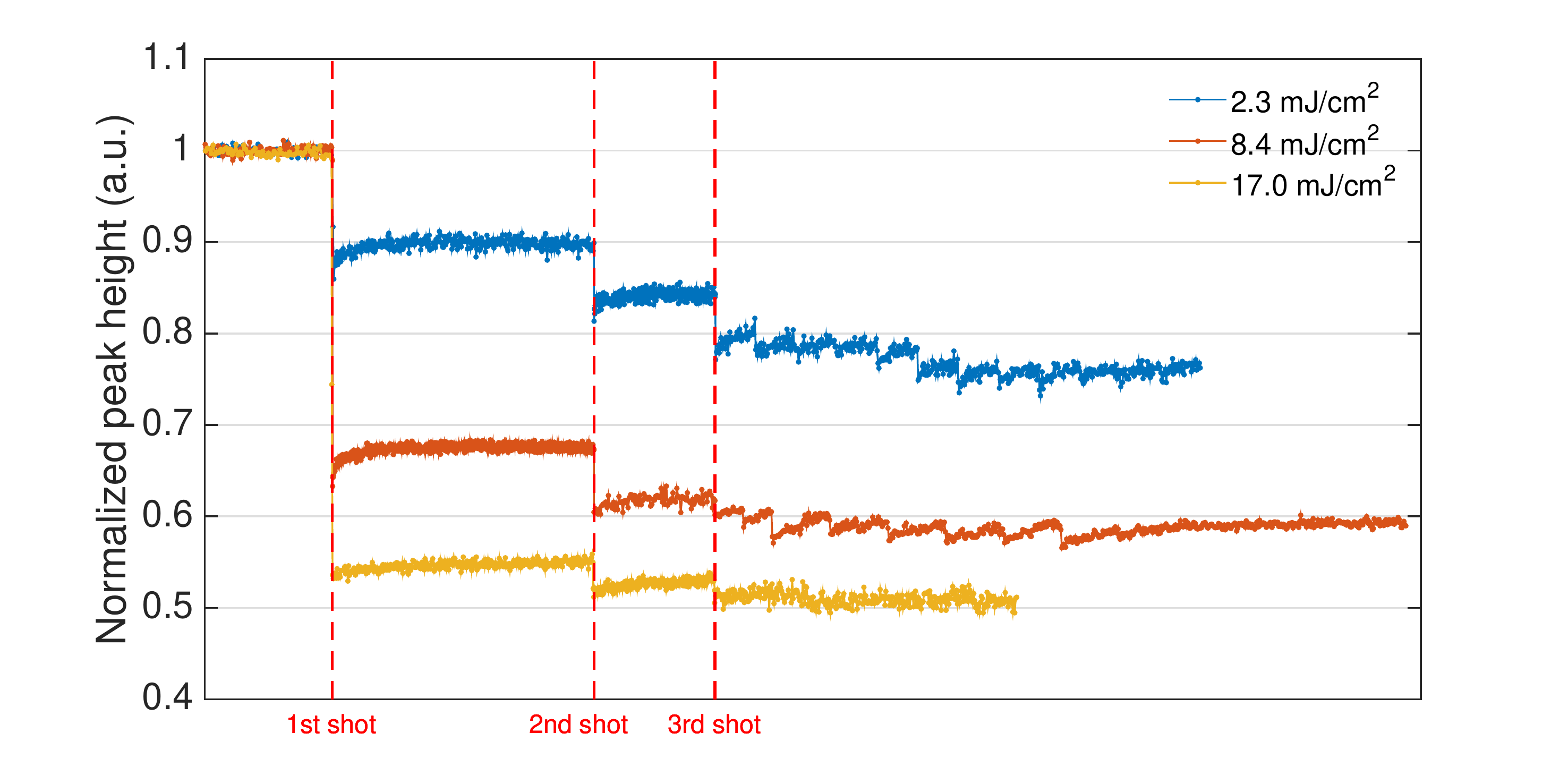}
\renewcommand{\figurename}{Fig. S}
\captionsetup{justification=raggedright,singlelinecheck=false}
\caption{\label{fig:wide}\textbf{Evolution of the magnetic peak height under excitation of multiple shots:} The pristine magnetic order was degraded by a sequence of single laser shots, the first three single shots were marked by red dashed lines. With increasing the laser fluence, the degree of the suppression of magnetic order keep reducing in the first initial stage where permanent suppression occurs; At high fluence, the multiple shots after the very first single shot drives marginal further suppression to the magnetic order and the system enters into a stable stage. }
\label{fig:multishots}
\end{figure*}

Magnetic peak height was measured upon a sequence of laser shots of various fluences. As shown in \textbf{Fig. S\ref{fig:multishots}}, each of the first a few shots induced certain degree of suppression to the magnetic peak height. After those initial shots, the peak height does fully recover, but only to a reduced level prepared by the initial multiple pulses. The stabilized conditions are obviously dependent on the laser pulse fluence.


\section{\label{sec:level2}Modeling the Thickness Dependence of the AFM Ordering at Finite Temperature}

\begin{figure}[htb]
\includegraphics[width=0.45\textwidth]{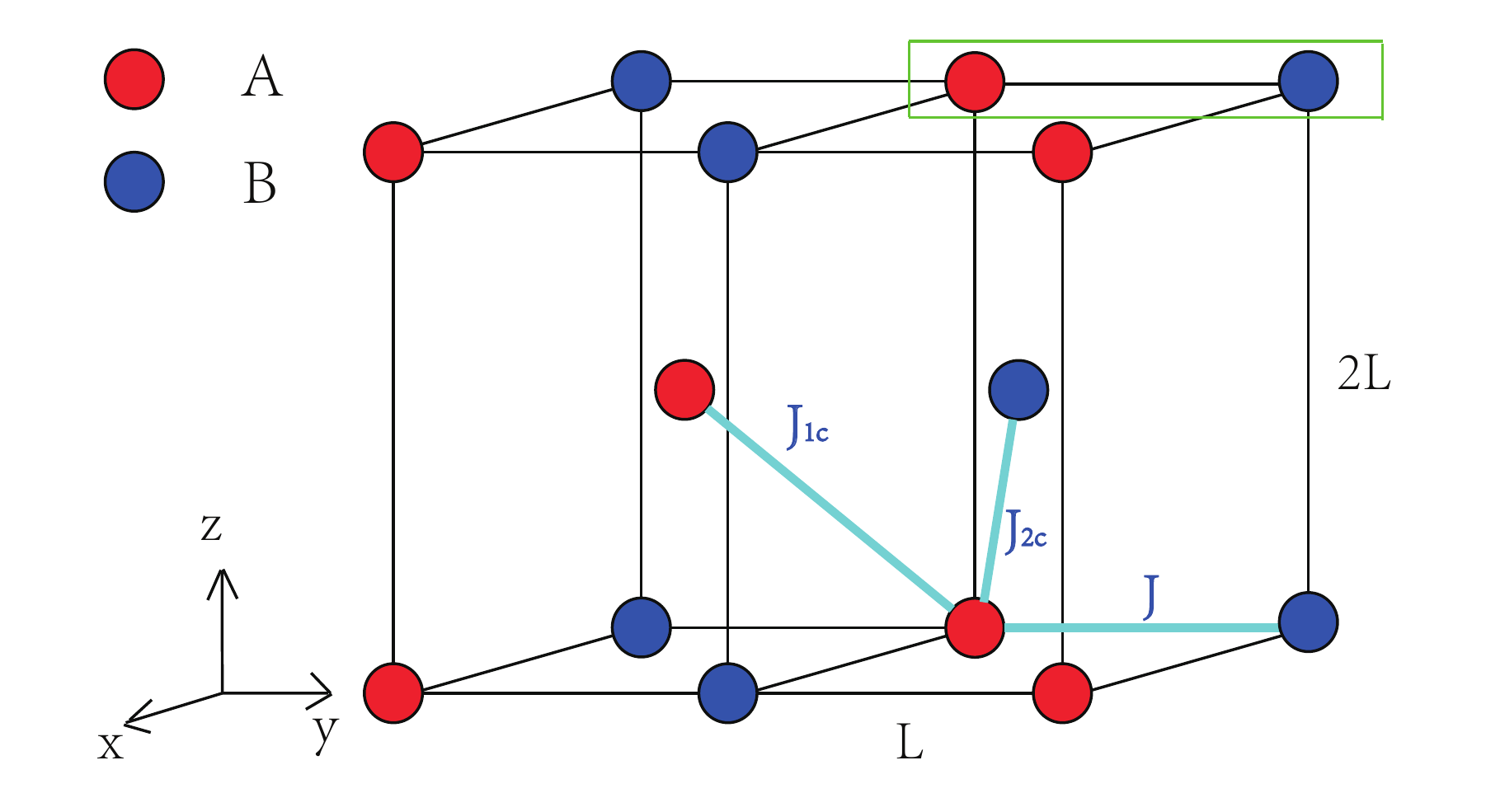}
\renewcommand{\figurename}{Fig. S}
\captionsetup{justification=raggedright,singlelinecheck=false}
\caption{\label{fig:wide}\textbf{Minimum model of an AFM ordered spin-1/2 system:} The ordered spins are grouped into the spin-up (A) and spin-down (B) sub-lattices. The exchange terms considered are labeled accordingly.}
\label{fig:theo_struc}
\end{figure}

We evaluate the saturation of the low limit of the observed inter-plane correlation by considering a minimum Quasi-2D spin model. As shown in \textbf{Fig. S\ref{fig:theo_struc}}, spin-$1/2$ objects are placed at the Ir sites which are AFM ordered, forming spin-up (A) and spin-down (B) sublattices. The exchange interactions considered are: the nearest-neighbor in-plane AFM exchange $J$, the inter-plane next-nearest-neighbor exchange $J_{1c}$ and $J_{2c}$ for the coupling within and between the spin-up and spin-down sublattices. Also, the anisotropy of the nearest-neighbor exchange, $\Delta$, is considered. As a result, the Hamiltonian for this minimum spin model can be written as\cite{diep1991, Moschel1993}, 

\begin{widetext}
\begin{eqnarray}
\begin{split}
H & = \frac{J}{2}\sum_{l<ij>}\vec{S_{li}}\cdot \vec{S_{lj}}
          + \Delta \sum_{l<ij>}S^z_{li}S^z_{lj}
     +\frac{J_c}{2}\sum_{<ll'><ij>}\vec{S_{li}}\cdot \vec{S_{l'j}}\\
  &=\frac{J}{2}[D\sum_{l<ij>}S^z_{lAi}S^z_{lBj}+\frac{1}{2}\sum_{l<ij>}(S^+_{lAi}S^-_{lBj}+S^-_{lAi}S^+_{lBj})]\\
      &+J[D\sum_{l<ij>}S^z_{lBi}S^z_{lAj}+\frac{1}{2}\sum_{l<ij>}(S^+_{lBi}S^-_{lAj}+S^-_{lBi}S^+_{lAj})]\\
     &+\frac{J_{1c}}{2}[\sum_{<ll'>}\sum_{<ij>}S^z_{lAi}S^z_{l'Aj}
       +\frac{1}{2}\sum_{<ll'>}\sum_{<ij>}(S^+_{lAi}S^-_{l'Aj}+S^-_{lAi}S^+_{l'Aj})]\\
    &+\frac{J_{1c}}{2}[\sum_{<ll'>}\sum_{<ij>}S^z_{lBi}S^z_{l'Bj}
       +\frac{1}{2}\sum_{<ll'>}\sum_{<ij>}(S^+_{lBi}S^-_{l'Bj}+S^-_{lBi}S^+_{l'Bj})]\\
    &+\frac{J_{2c}}{2}[\sum_{<ll'>}\sum_{<ij>}S^z_{lAi}S^z_{l'Bj}
       +\frac{1}{2}\sum_{<ll'>}\sum_{<ij>}(S^+_{lAi}S^-_{l'Bj}+S^-_{lAi}S^+_{l'Bj})]\\
    &+\frac{J_{2c}}{2}[\sum_{<ll'>}\sum_{<ij>}S^z_{lBi}S^z_{l'Aj}
       +\frac{1}{2}\sum_{<ll'>}\sum_{<ij>}(S^+_{lBi}S^-_{l'Aj}+S^-_{lBi}S^+_{l'Aj})]\\
\end{split}
\end{eqnarray}  
\end{widetext}
where $D$ is defined as $D = 1 + \Delta $.

To compare with our experimental observations on \SIO{}, we refer to the published literature\cite{Porras2019, YGim2016, Calder2018} and set, 


\begin{itemize}
\item Nearest neighbor exchange interaction $J = 60~meV$
\item Interlayer exchange interaction: within the same sublattice $J_{1c}=-16.4~\mu eV$; between the two sublattices $J_{2c}=16.4~\mu eV$
\item the anisotropic term related to the magnon gap as $E(k) = ZJ\langle Sz \rangle \sqrt{D^2-\gamma_k^2}$
\end{itemize}

As the reported exchange anisotropy for Sr2IrO4 is quite controversial\cite{Porras2019, YGim2016, Calder2018, Pincini2017, Bahr2014}, the gap size was carried as a free parameter in our calculation. The dynamics of the above Hamiltonian was solved with the equation of motion
technique and mean-field approximation for model systems with different thickness.

Using the equation of motion of double-time Green's function\cite{diep1991, Moschel1993, Kuntz2006}, we can obtain
a set of Green's functions for each layer: 

\begin{widetext}
\begin{eqnarray}
\begin{split}
&[\omega-JDZ\langle S^{z}_{l} \rangle +Z'(J_{1c}-J_{2c})(\langle S^{z}_{l+1} \rangle + \langle S^{z}_{l-1} \rangle)]g_{ll}
    -J\langle S^{z}_{l} \rangle Z \gamma(k) f_{ll}\\
    &-J_{1c} Z' \gamma_{AA}(k)(g_{l-1,l}(k)+g_{l+1,l}(k))-J_{2c} Z' \gamma_{AB}(k)(f_{l-1,l}(k)+f_{l+1,l}(k))=2\langle S^{z}_{l} \rangle\\
&\\
& [\omega + JDZ\langle S^{z}_{l} \rangle  - Z'(J_{1c}-J_{2c})(\langle S^{z}_{l+1} \rangle + \langle S^{z}_{l-1} \rangle)]f_{ll}
    +J\langle S^{z}_{l} \rangle Z \gamma(k) g_{ll}\\
    &+J_{1c} Z' \gamma_{AA}(k)(f_{l-1,l}(k)+f_{l+1,l}(k)) + J_{2c} Z' \gamma_{AB}(k)(g_{l-1,l}(k)+g_{l+1,l}(k))=0 \\
&\\
&[\omega-JDZ\langle S^{z}_{l-1} \rangle +Z'(J_{1c}-J_{2c})(\langle S^{z}_{l} \rangle + \langle S^{z}_{l-2} \rangle)]g_{l-1,l}
    -J\langle S^{z}_{l-1} \rangle Z \gamma(k) f_{l-1,l}\\
    &-J_{1c} Z' \gamma_{AA}(k)\langle S^{z}_{l-1} \rangle(g_{l-2,l}(k)+g_{l,l}(k))
      -J_{2c} Z' \gamma_{AB}(k)\langle S^{z}_{l-1} \rangle(f_{l-2,l}(k)+f_{l,l}(k))=0\\
&\\
& [\omega + JDZ\langle S^{z}_{l-1} \rangle  - Z'(J_{1c}-J_{2c})(\langle S^{z}_{l-2} \rangle + \langle S^{z}_{l} \rangle)]f_{l-1,l}
    +J\langle S^{z}_{l-1} \rangle Z' \gamma(k) g_{l-1,l}\\
    &+J_{1c} Z' \gamma_{AA}(k)\langle S^{z}_{l-1} \rangle(f_{l-2,l}(k)+f_{l,l}(k))
     + J_{2c} Z' \gamma_{AB}(k)\langle S^{z}_{l-1} \rangle(g_{l,l-2}(k)+g_{l,l}(k))=0 \\
&\\
\end{split}
\end{eqnarray}
\end{widetext}

Where N is the total number of layers, and $l$ is the index of each layer ($l = 1,2,...,N$). $Z, Z'$ are the in-plane and out of plane coordinate numbers. $g_{ll'}(k)$ and $f_{ll'}(k)$ are Fourier transformation of $G_{ll'}(\omega) $
 and $F_{ll'}(\omega)$ in k-space. $\gamma$'s are the geometry factors: 
\begin{itemize}
\item $\gamma(k)$: in-plane between A and B Ir atoms
\item $ \gamma_{AA}(k)$: Nearest  layers between A-A or B-B Ir atoms
\item $ \gamma_{AB}(k)$: Nearest layers between A-B Ir atoms
\end{itemize}

The equation of motion was solved self-consistently for T = 80 K, which is our experimental temperature. Once the local correlation function,  
\begin{equation}
\langle  S^{-}_{l} S^{+}_{l}   \rangle = \frac{i}{2\pi} \int^{\infty}_{-\infty} 
           \frac{d\omega}{e^{\frac{\omega}{k_B T}}+1}
           \{ g_{ll}(\omega+i0^{+})-g_{ll}(\omega - i0^{-})\}
\end{equation}
is obtained from the closed self-consistent loop, the local magnetic moments
$\langle S^{z}_{l} \rangle$ of each layer, 
\begin{equation}
  \langle S^{z}_l \rangle = \frac{1}{2} - \langle S^{-}_l S^{+}_l \rangle
\end{equation}
was extracted and plotted as \textbf{Fig. 4} in the main text.
